\documentclass[english]{bvm2018}
\usepackage{hyperref}

\title{An open-source tool for automated planning of overlapping ablation zones for percutaneous renal tumor treatment}

\titlerunning{Automated planning of overlapping ablation zones}

\author{A. M. Franz$^{1}$, B. J. Mittmann$^1$, J. Roeser$^2$, B. Schmidberger$^2$, M. Meinke$^3$, P.~L.~Pereira$^4$, H. U. Kauczor$^3$, G. M. Richter$^5$  and C. M. Sommer$^{3,5}$}

\authorrunning{Franz et al.}
\institute{
$^1$Institute for Computer Science, Ulm University of Applied Sciences, Ulm, Germany\\
$^2$Institute for Medical Engineering and Mechatronics, University of Applied Sciences, Ulm, Germany\\
$^3$Clinic of Diagnostic and Interventional Radiology, University Hospital Heidelberg, Heidelberg, Germany\\
$^4$Clinic of Radiology, Minimally-invasive Therapies and Nuclear Medicine, SLK Kliniken Heilbronn GmbH, Heilbronn, Germany \\
$^5$Clinic of Diagnostic and Interventional Radiology, Stuttgart Clinics, Katharinenhospital, Stuttgart, Germany \\
}

\email{franz@hs-ulm.de}

\begin{document}

%==============================================================================
% w�hlen Sie mit dem Befehl \selectlanguage die Sprache aus, in der Ihr 
% Proceeding verfasst ist
%
%\selectlanguage{german}
\selectlanguage{english}

\maketitle

\begin{abstract}
Percutaneous thermal ablation is a minimally-invasive treatment option for renal cancer. To treat larger tumours, multiple overlapping ablations zones are required. Arrangements with a low number of ablation zones but coverage of the whole tumour volume are challenging to find for physicians. In this work, an open-source software tool with a new planning approach based on the automatic selection from a large number of randomized geometrical arrangements is presented. Two uncertainty parameters are introduced to account for tissue shrinking and tolerance of non-ablated tumour volume. For seven clinical renal T1a, T1b and T2a tumours, ablation plans were proposed by the software. All proposals are comparable to manual plans of an experienced physician with regard to the number of required ablation zones.
\end{abstract} 

\section{Introduction}

More than 14.000 patients are diagnosed with renal cancer per year in Germany. Besides surgical treatment, percutaneous ablation was established as a minimally-invasive alternative \cite{Filippiadis2019}. Therefore, a needle-shaped ablation probe is inserted into the affected tissue, commonly under computed-tomography (CT) or magnetic resonance imaging (MRI) guidance. The tumour is then destroyed thermically, e.g., by using radiofrequency or microwaves. The whole tumour must be ablated for a successful treatment. Researchers proposed methods to estimate the treated area of ablations \cite{Voglreiter2018,Pena2019}. For example, vessel information from preoperative data can be used to compute cooling effects \cite{Huang2013}, but segmenting a vessel tree often is too cumbersome in clinical routine. Instead, physicians commonly estimate a spherical or elliptical shape of a single ablation, referred to as ablation zone. These shapes are given by manufacturers for their probes depending on the ablation duration and the applied power. Usually, a security margin around the tumour is ablated to account - among other things - for uncertainties in the estimation of ablation zones. If a small tumour can be completely covered by a single ablation, the probe is inserted to the center. For larger tumours, e.g. with a maximum diameter of 4~cm (T1a for renal tumours) or 7~cm (T1b), multiple overlapping ablations zones (MOAZs) are required. Manually finding a geometric distribution that covers the whole volume in 3D with a small number of ablation zones is challenging. Algorithms for planning MOAZs can use ideal geometric combinations of spheres to fully cover a volume\cite{Dodd2001,Yang2010}. However, this leads to relatively high numbers of ablations for larger tumours. Planning MOAZs can also be treated as an optimization problem, as proposed by Ren et al., who used a branch and bound algorithm to solve planning tasks for tumours with diameters of 2.5~cm and 3.5~cm simulated in a porcine model \cite{Ren2014}. % and found 2-3 respectively 6-7 required ablations \cite{Ren2014}.

%\cite{Chen2007} TODO

In this work we present an open-source software tool for a planning approach based on the automatic selection from a large number of randomized geometrical arrangements under consideration of uncertainty parameters that can be adjusted to get plans with an acceptable number of ablations. It was evaluated on renal tumours in comparison to manual plans of a physician.

%\cite{Baegert2007} trajectory planning (weglassen?)

\section{Materials and Methods}

For automated planning of MOAZs, the following workflow steps are proposed: (I) segmentation of the tumour in preoperative images, (II) definition of safety margin, (III) definition of diameter for spherical ablation zones, (IV) definition of uncertainty parameters for ablation zone placement, (V) definition of a model for geometrical distribution and (VI)  computation of MOAZs.

Segmentation of the tumour in step I was performed manually using the Medical Imaging Interaction Toolkit (MITK)\cite{Nolden2012}. In the following, the uncertainty parameters for step IV are described and a model based on the automatic selection from a large number of randomized geometrical arrangements for step V is presented. Finally, the open-source implementation and experiments including example parameters for step II and III are explained.

\subsection{Definition of uncertainty parameters}
Estimating the ablated tissue volume during a thermal ablation is usually subject to uncertainties such as tissue shrinkage after ablations \cite{Farina2018} or other effects. Due to their experience, physicians account for these uncertainties by adapting their manual ablation plans during conventional treatment. This includes adaptation of the ablation diameter because reference values are given for the state after tissue shrinking and also tolerance of small areas of tumour tissue which is not directly treated. To account for this during automated planning, two Ulm-Heidelberg-Uncertainty-parameters (UHUs) are proposed: 
\begin{itemize}
	\item UHU-1: Tolerance of non-ablated tumour volume [\% of tumour volume]
	\item UHU-2: Tissue shrinkage after ablation [\% of ablated volume]
\end{itemize}

\subsection{Approach for planning overlapping ablation zones}
\begin{figure}[t]
	\centering
	\includegraphics[width=1.0\textwidth]{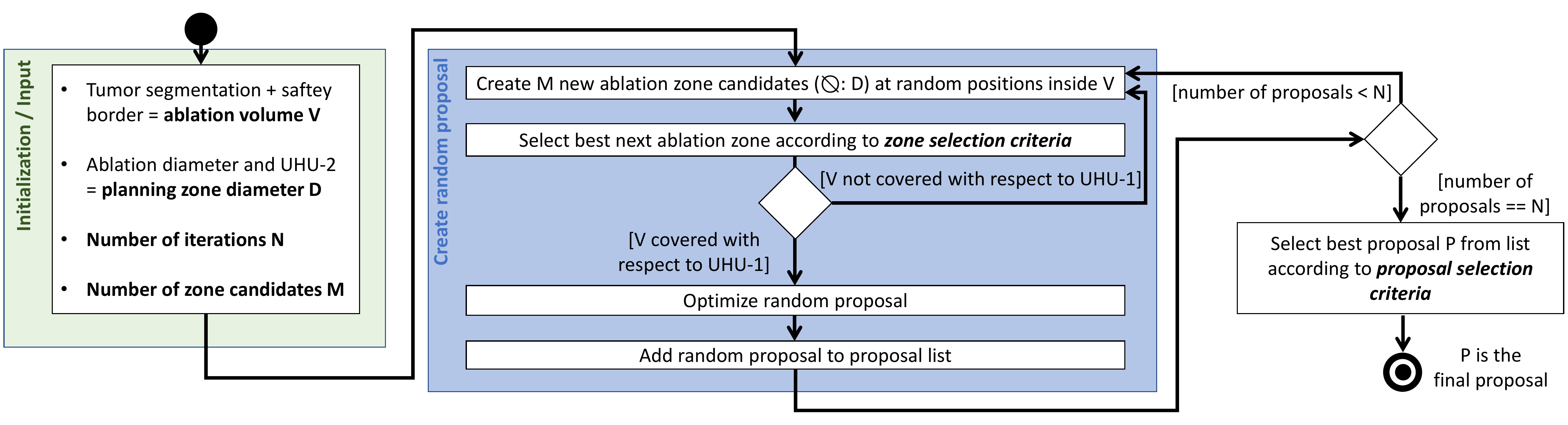}
	\caption{Approach for automated proposal of overlapping ablation zones.} 
	\label{fig:approach}
\end{figure}

The proposed approach is shown in Fig.~\ref{fig:approach}. For initialization, a safety margin (in mm) is added to the tumour segmentation which increases the area to be ablated to volume~V. The ablation zone diameter represents the maximum reachable diameter of a single ablation. It is increased by the factor given from UHU-2 resulting in the planning zone diameter D. Further, the number of iterations N and the number of zone candidates M are defined.

A single random proposal is created by adding new zones as long as V is not covered with respect to UHU-1. A new zone is selected from M candidates according to zone selection criteria which can include coverage of new tumour volume and percentage of overlap with other zones or non-tumour tissue. If enough zones are found, the proposal is optimized by moving zones that overlap to non-tumour tissue too much towards the center of the tumour, removing zones that overlap too much with other zones and/or decreasing the diameter of single ablations if possible without loosing coverage of V.

After iteratively adding N proposals to a list, the best proposal P is selected according to the proposal selection criteria. The most important criterion typically is a low number of required ablation zones. %Secondary criteria can be the percentage of overlap between ablation zones or to non-tumour tissue.

\subsection{Open-source implementation}
An open-source plugin for MITK was implemented to test the approach. It supports selecting a segmentation and adding a safety margin. The parameters N, UHU-1, UHU-2 and the zone diameter can be defined before a planning proposal is computed. The other parameters and criteria are fixed in the implementation: Parameter M is set to 1 for the starting zone and to 5 for all following zones. The overlap of a zone and V is used as zone selection criterion, while always the zone with the highest overlap is chosen. The proposal with the lowest number of required ablations is selected as best proposal. %Segmentation features are supported by existing MITK plugins \cite{Nolden2012}.

In the optimization step for each proposal, zones that overlap more than 30\% into tissue outside V are moved towards the center of the tumour until less than 30\% are outside. Afterwards, ablation zones are reduced in diameter and/or ablation zones are removed if possible without loosing coverage of V.

\begin{figure}[t]
	\centering
	\includegraphics[width=1.0\textwidth]{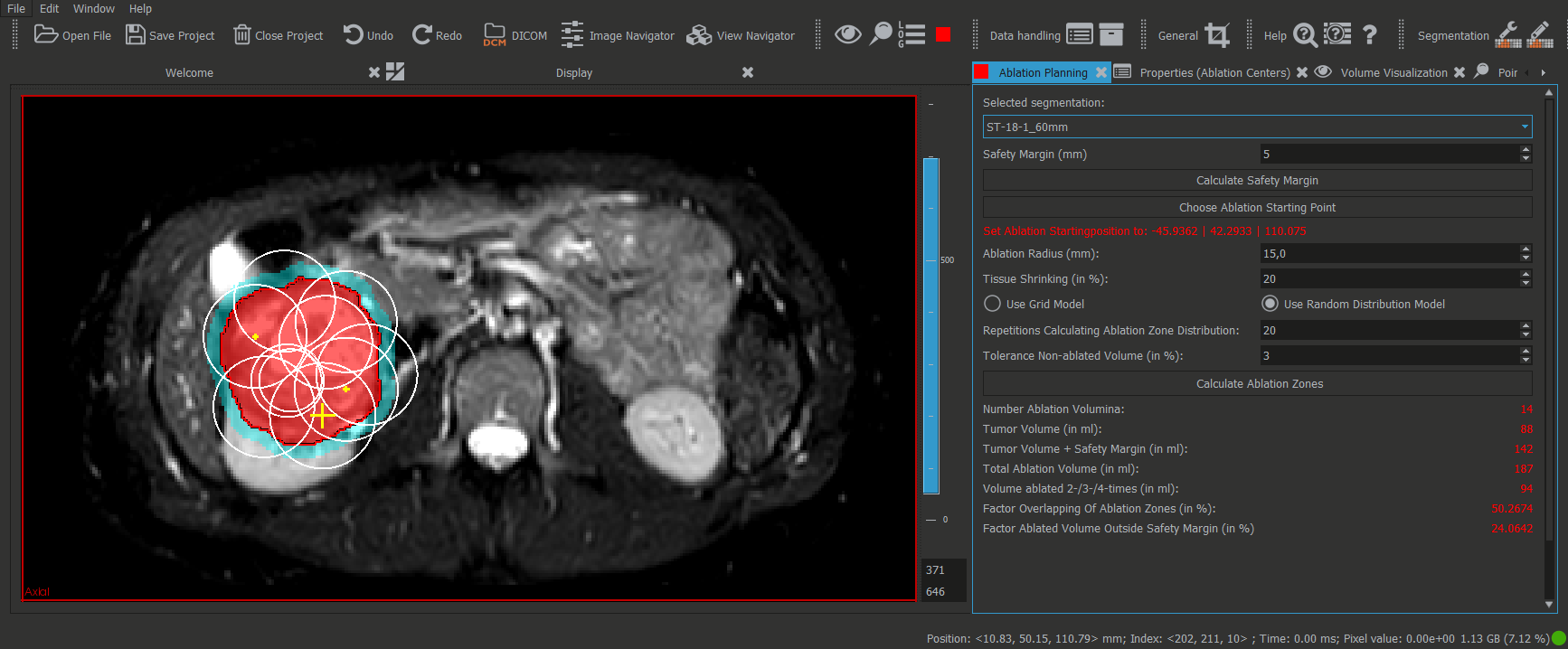}
	\caption{Computed planning proposal for a tumour (red) segmented in a MRI image with added safety margin (blue) and ablation zones (white circles).} 
	\label{fig:screenshot}
\end{figure}

For the final proposal P, ablation spheres are visualized and statistical data is reported, including the number of ablation zones, the tumour volume with and without safety margin and the ablated volume. Fig.~\ref{fig:screenshot} shows a screenshot of the plugin and a computed planning proposal.

\subsection{Experiments}

For the two experiments, CT data of renal tumours from four publicly available datasets (\mbox{C3N-00312}, \mbox{C3N-00305}, \mbox{C3N-00491} and \mbox{TCGA-KM-8442} from collections CPTAC-CCRCC and TCGA-KICH, \href{http://www.cancerimagingarchive.net}{http://www.cancerimagingarchive.net}) and three anonymized clinical cases from the University Hospital Heidelberg and Stuttgart were used. A zone diameter of 3~cm was chosen which is a typical maximum diameter of spherical ablation zones, e.g. of the Emprint Ablation System (Medtronic plc., Dublin, Ireland) for renal ablations (75 watts / 5:30 minutes duration). 50 iterations were chosen because this leads to acceptable computational time of less than 5 minutes for T1a an T1b tumours and reproducible results concerning the number of required zones. All tumours were manually segmented using existing MITK plugins \cite{Nolden2012} and a safety margin of 5 mm was added. 

In experiment 1, a 50 mm T1b tumour (C3N-00305) was used to test the influence of UHU-1 and UHU-2 to the planning proposals. The parameters were chosen as follows: UHU-1: 0/3/6/10/20\%; UHU-2: 0/10/20/30\%. For each parameter configuration, three ablation plans were automatically proposed by the software prototype. The plans were evaluated regarding the number of ablation zones and also manually examined concerning a meaningful zone distribution. %and the ablation volume outside tumour and safety margin [\%]

In experiment 2, all seven datasets were used to compare the automated planning proposals with manual plans of an experienced physician. For each tumour, three automated planning proposals were computed with UHU-1 of 3\% and UHU-2 of 20\%. The number of required ablations is reported for each automated planning proposal as well as for the independent manual plan.

\section{Results}

\begin{table}
	\centering
  \begin{tabular}{ c | c | c | c | c | c }
		Data set & Tumour \O [mm] & Manual plan & Proposal 1 & Proposal 2 & Proposal 3 \\
    \hline
    C3N-00312 & 40 (T1a) & 5 & 6 & 5 & 6 \\ \hline
    C3N-00305 & 50 (T1b) & 8 & 5 & 4 & 5 \\ \hline
    C3N-00491 & 45 (T1b) & 10 & 13 & 13 & 13 \\  \hline
		Anonym.1 & 40 (T1a) & 5 & 7 & 7 & 7 \\  \hline
		Anonym.2 & 60 (T1b) & 14 & 13 & 14 & 13 \\  \hline
		Anonym.3 & 40 (T1a) & 7 & 7 & 7 & 6 \\  \hline
		TCGA-KM-8442 & >70 (T2a) & 18 & 20 & 22 & 21 \\ \hline
    \hline
  \end{tabular}
\caption{Number of ablation zones (\O:30 mm) of a manual plan of an experienced physician and three automatic proposals (UHU-1: 3\%; UHU-2: 20\%)}
    \label{tab:resultsExp2}   
\end{table}

An installer of the software prototype will be provided in the Open Science Framework (page \href{https://osf.io/r7f5d/}{https://osf.io/r7f5d/}) together with segmentations of the four publicly available data sets upon publication of this work. For the T1a and T1b tumours, automatic planning required up to 5 minutes (PC: core i7, 24 GB Ram). The T2b tumour required around 60 minutes.

For data set C3N-00305 with varying UHU-1 and UHU-2, the number of required MOAZs ranged from $2\pm0$ ($\mu\pm\sigma$, n=3, UHU-1:20\%, UHU-2:30\%) to $11\pm0$ (n=3, UHU-1:0\%, UHU-2:0\%). Manually examining the ablation plans, we found that too high uncertainty lead to plans that don't cover the whole tumour while low or no uncertainty leads to a too large number of ablations. Parameters of UHU-1:3\% and UHU-2:20\% lead to $5\pm0$ (n=3) MOAZs, which a physician confirmed to be a good trade-off for a realistic ablation plan.

The amount of required MOAZs for different tumours are shown in Table~\ref{tab:resultsExp2}. Manual plans of an experienced physician required 5/5/7 (T1a tumours), 8/10/14 (T1b) and 18 (T2a) zones, while the automatic planning with UHU-1: 3\% and UHU-2: 20\% lead to 5.7/7.0/6.7 (T1a), 4.7,13.0,13.7 (T1b) and 21.0~(T2a) zones on average (n=3).

\section{Discussion}

The experiments confirmed, that planning MOAZs based on an automatic selection from a large number of randomized geometrical arrangements and uncertainty parameters is feasible in acceptable time for T1a and T1b tumours (less than 5 minutes). Testing the software prototype with seven clinical data sets lead to results comparable to manual plans of a physician (c.f. Table~\ref{tab:resultsExp2}). The software was able to propose a plan for a T2a tumour, while computational time raised to around 60 minutes. Such large tumours are usually not treated with ablation therapy because it would require too much time to perform 20 or more ablations. However, ongoing studies show that even this is feasible \cite{Schullian2019}.

In a first try, we implemented a model based on an ideal grid of overlapping spheres, similar to Yang et al. (c.f. Fig.~6 in \cite{Yang2010}). Looking at the results, we found a too large number of required MOAZs in all cases (e.g., over 20 for C3N-00491). 

Our approach incorporates uncertainty parameters that are subject to ongoing clinical research \cite{Farina2018}. It needs to be shown, that a certain parameter set, such as 20\% of tissue shrinking and 3\% tolerance of non-ablated tumour volume as proposed in this work, leads to clinical acceptable outcome. A first step in this direction might be an ex-vivo or in-vivo animal study. Further, only spherical MOAZs are supported so far. While this might be sufficient for some ablation probes, future work includes an extension to elliptical ablation zones.

To improve computational time, it is planned to parallelize the implementation of the approach shown in Fig.~\ref{fig:approach} which is expected to be straight-forward, because the creation of N proposals can be done in parallel. It remains to be discussed if an algorithm highly based on random components would be acceptable in clinics, which leads to general ethical questions that also raise up for other more advanced methods, such as machine learning. 

%Maybe: Discuss segmentation
%Maybe: Future changes on the software: other zone selection and proposal selection criteria

%\section*{Acknowledgements}
%******
%This work was supported by the European Union through the ERC starting Grant COMBIOSCOPY (ERC-2015-StG-37960) and by the Irish Health Research Board (POR/2012/31), Science Foundation Ireland (15/TIDA/2846). We acknowledge the German Cancer Research Center (DKFZ), Heidelberg, and the Institute of Image-Guided Surgery (IHU), Strasbourg for supporting this work.

\bibliographystyle{bvm2019}

\bibliography{AblationPlanning}

\begin{thebibliography}{10}

\bibitem{Filippiadis2019}
Filippiadis D, Mauri G, Marra P, Charalampopoulos G, Gennaro N, De~Cobelli F.
\newblock {{P}ercutaneous ablation techniques for renal cell carcinoma: current
  status and future trends}.
\newblock Int J Hyperthermia. 2019;36(2):21--30.

\bibitem{Voglreiter2018}
Voglreiter P, Mariappan P, Pollari M, Flanagan R, Blanco~Sequeiros R,
  Portugaller RH, et~al.
\newblock RFA Guardian: Comprehensive Simulation of Radiofrequency Ablation
  Treatment of Liver Tumors.
\newblock Nature Scientific Reports. 2018;8(1):787.

\bibitem{Pena2019}
Pena K, Ishahak M, Arechavala S, Leveillee RJ, Salas N.
\newblock {{C}omparison of temperature change and resulting ablation size
  induced by a 902-928 {M}{H}z and a 2450 {M}{H}z microwave ablation system in
  in-vivo porcine kidneys}.
\newblock Int J Hyperthermia. 2019;36(1):313--321.

\bibitem{Huang2013}
Huang HW.
\newblock {{I}nfluence of blood vessel on the thermal lesion formation during
  radiofrequency ablation for liver tumors}.
\newblock Med Phys. 2013;40(7):073303.

\bibitem{Dodd2001}
Dodd GD, Frank MS, Aribandi M, Chopra S, Chintapalli KN.
\newblock {{R}adiofrequency thermal ablation: computer analysis of the size of
  the thermal injury created by overlapping ablations}.
\newblock AJR Am J Roentgenol. 2001 Oct;177(4):777--782.

\bibitem{Yang2010}
{Yang} L, {Wen} R, {Qin} J, {Chui} C, {Lim} K, {Chang} SK.
\newblock A Robotic System for Overlapping Radiofrequency Ablation in Large
  Tumor Treatment.
\newblock IEEE/ASME Transactions on Mechatronics. 2010;15(6):887--897.

\bibitem{Ren2014}
Ren H, Campos-Nanez E, Yaniv Z, Banovac F, Abeledo H, Hata N, et~al.
\newblock {{T}reatment planning and image guidance for radiofrequency ablation
  of large tumors}.
\newblock IEEE J Biomed Health Inform. 2014;18(3):920--928.

\bibitem{Nolden2012}
Nolden M, Zelzer S, Seitel A, Wald D, Muller M, Franz AM, et~al.
\newblock {{T}he {M}edical {I}maging {I}nteraction {T}oolkit: challenges and
  advances : 10 years of open-source development}.
\newblock Int J Comput Assist Radiol Surg. 2013;8(4):607--620.

\bibitem{Farina2018}
Farina L, Nissenbaum Y, Cavagnaro M, Goldberg SN.
\newblock {{T}issue shrinkage in microwave thermal ablation: comparison of
  three commercial devices}.
\newblock Int J Hyperthermia. 2018 06;34(4):382--391.

\bibitem{Schullian2019}
Schullian P, Johnston EW, Putzer D, Eberle G, Laimer G, Bale R.
\newblock {{S}tereotactic radiofrequency ablation of subcardiac hepatocellular
  carcinoma: a case-control study}.
\newblock Int J Hyperthermia. 2019;36(1):876--885.

\end{thebibliography}
% Bitte setzen Sie hier Ihre Beitragsnummer ein und benennen Sie
% die BibTeX-Datei ebenfalls auf Ihre Beitragsnummer um.
%Kontrollzeiledef
\end{document}